\documentclass[%
superscriptaddress,
onecolumn,
nofootinbib,
 amsmath,amssymb,
 aps,
prb,
]{revtex4-2}

\usepackage{graphicx}
\usepackage[T1]{fontenc}
\usepackage[utf8]{inputenc}
\usepackage[UKenglish]{babel}
\usepackage{placeins}
\usepackage[binary-units=true]{siunitx}
\usepackage{color}
\usepackage[dvipsnames]{xcolor}
\usepackage{nicefrac}
\usepackage{mathtools}
\usepackage{bbold}
\usepackage{braket}
\usepackage{xspace}
\usepackage{feynmp-auto}
\usepackage[
colorlinks=true,
allcolors=blue
]{hyperref}
\usepackage{cleveref}

\graphicspath{{figures/},{inkscape/},{data/}}

\DeclareMathOperator{\im}{i}

\newcommand{\pdagger}{{\phantom{\dagger}}}
\newcommand{\matr}[2]{\left(\begin{array}{#1}#2\end{array}\right)}

\newcommand{\eto}[1]{\ensuremath{\mathrm{e}^{#1}}}

\newcommand{\fuj}{Fuji\xspace}
\newcommand{\kil}{Kilimanjaro\xspace}

\begin{document}
\title{An Effective Theory for Graphene Nanoribbons with Junctions}
	
	\author{Johann Ostmeyer}%
	\email{j.ostmeyer@liverpool.ac.uk}
	\affiliation{Department of Mathematical Sciences, University of Liverpool, Liverpool, L69 7ZL, United Kingdom}
	
	\author{Lado Razmadze}
	\email{l.razmadze@fz-juelich.de}
	\affiliation{Institute for Advanced Simulation (IAS-4), Forschungszentrum J\"ulich, Germany}
	
	\author{Evan Berkowitz}
	\email{e.berkowitz@fz-juelich.de}
	\affiliation{Institute for Advanced Simulation (IAS-4), Forschungszentrum J\"ulich, Germany}
	\affiliation{J\"{u}lich Supercomputing Center, Forschungszentrum J\"{u}lich, 54245 J\"{u}lich, Germany}
	\affiliation{Center for Advanced Simulation and Analytics (CASA), Forschungszentrum Jülich, 52425 J\"{u}lich, Germany}
	
	\author{Thomas Luu}%
	\email{t.luu@fz-juelich.de}
	\affiliation{Institute for Advanced Simulation (IAS-4), Forschungszentrum J\"ulich, Germany}
	\affiliation{Helmholtz-Institut f\"ur Strahlen- und Kernphysik and Bethe Center for Theoretical Physics, Rheinische Friedrich-Wilhelms-Universit\"at Bonn, Germany}

	\author{Ulf-G. Mei{\ss}ner}
	\email{meissner@hiskp.uni-bonn.de}
	\affiliation{Helmholtz-Institut f\"ur Strahlen- und Kernphysik and Bethe Center for Theoretical Physics, Rheinische Friedrich-Wilhelms-Universit\"at Bonn, Germany}
	\affiliation{Institute for Advanced Simulation (IAS-4), Forschungszentrum J\"ulich, Germany}
	\affiliation{Center for Advanced Simulation and Analytics (CASA), Forschungszentrum Jülich, 52425 J\"{u}lich, Germany}
	\affiliation{Tbilisi State University, 0186 Tbilisi, Georgia}

	\date{\today}
	\begin{abstract}
		Graphene nanoribbons are a promising candidate for fault-tolerant quantum electronics.
		In this scenario, qubits are realised by localised states that can emerge on junctions in hybrid ribbons formed by two armchair nanoribbons of different widths.
		We derive an effective theory based on a tight-binding ansatz for the description of hybrid nanoribbons and use it to make accurate predictions of the energy gap and nature of the localisation in various hybrid nanoribbon geometries.
		We use quantum Monte Carlo simulations to demonstrate that the effective theory remains applicable in the presence of Hubbard interactions.
		We discover, in addition to the well known localisations on junctions, which we call `\fuj', a new type of `\kil' localisation smeared out over a segment of the hybrid ribbon.
		We show that \fuj localisations in hybrids of width $N$ and $N+2$ armchair nanoribbons occur around symmetric junctions if and only if $N\pmod3=1$, while edge-aligned junctions never support strong localisation.
		This behaviour cannot be explained relying purely on the topological $Z_2$ invariant, which has been believed the origin of the localisations to date.
	\end{abstract}
	\maketitle


\section{Introduction\label{sec:intro}}

The ability to engineer hybrid nanoribbons~\cite{rizzo18,groening2018} has opened up the possibility of using such systems to manufacture quantum dots~\cite{doi:10.1021/acsnano.1c09503} and other advanced electronic devices.
A central aspect that drives the usefulness of these systems is their ability to support localized electronic states that can be achieved through careful doping of the ribbons.
Various models of nanoribbons exhibit edge-state localization with a topological origin~\cite{Wakabayashi_2010,Ezawa2018,Yang:2020lal,lee2023mutual}.
In~\cite{ribbon_topology} it was argued that completely localized low-energy states occur at the junction of two armchair graphene nanoribbons (AGNRs) that are topologically distinct, forming so-called symmetry-protected topological edge states that should depend only on the geometrical, or topological, aspects of the system and not on the details of any interaction.
These states have electrons confined not only to the edge of the ribbon, but concentrated around the junctions.

Ref.~\cite{ribbon_junctions} confirmed that this localization is robust against the inclusion of an onsite Hubbard interaction via non-perturbative calculations.
The localization of states for the 7/9 and 13/15 hybrid nanoribbon systems persisted for a wide range of Hubbard interactions.
Recently the authors of Ref.~\cite{honet2023robust} have also investigated the role of interactions in ribbons with finite lengths using a mean-field prescription.
Other interesting phenomena occur when certain symmetries, such as the sublattice or \emph{chiral} symmetry, is broken in these systems~\cite{lee2023chiral}.

Though states in these hybrid systems demonstrate localization originating at junctions between different distinct AGNRs, the exact asymptotic behavior of these localized states has not been quantified.
As a function of distance from a junction wavefunctions may fall off exponentially (`strong localization') or with some power law (`weak localization').
This distinction has ramifications for the engineering requirements for manufacturing ribbons that support localization.
As we show in this paper, ribbon junctions that support wavefunctions with exponential decays on either side can be constructed such that they are nearly gapless under the tight-binding approximation.
Further, localization in this case can occur for a hybrid system with a single junction.

On the other hand, weak localization on either side of the ribbon junction cannot support a zero mode.
Using weak localization to concentrate a state along a ribbon segment requires ribbons with an even number of junctions.

These findings are easily understood through an effective theory (ET) of the hybrid ribbons in one dimension (1-D).
We show how to construct such a theory, and demonstrate how the parameters of this ET can be tuned to reproduce the low-energy spectrum of hybrid ribbons, even in the presence of non-perturbative interaction.
Once tuned, it is much simpler to use this theory to ascertain the behavior of the low-energy spectrum of these systems for different ribbon lengths. 
Indeed, we use this ET to make predictions on the specifications of hybrid ribbons that lead to a (nearly) gapless system.
We verify the predictions of our ET by comparing directly with calculations on the original hybrid systems.

Our paper is organized as follows.
In Sec.~\ref{sec:ribbons} we review ribbons of uniform width and their non-interacting dispersion relations; whether a given width is gapped or not controls how electronic states are localized around the junctions of hybrid ribbons, which we demonstrate in Sec.~\ref{sec:junctions}.
If a uniform ribbon is gapped the wavefunction decays exponentially on a segment of that width near a junction, while if the uniform ribbon is not gapped the wavefunction decays only with an inverse power law.
From this understanding we develop and test an effective one-dimensional tight-binding Hamiltonian with two hopping amplitudes in Sec.~\ref{sec:theory}.
We show how the effective hopping amplitudes depend on the specific geometries of the hybrid ribbons, identifying low-energy constants (LECs) that depend on the width of the ribbon segments but not on their lengths.
After fitting these LECs we demonstrate how our ET predicts ribbon widths and lengths that have a nearly-gapless spectrum.
We extend the validity of the ET to hybrid ribbons with Hubbard interaction by introducing an additional LEC and verify correctness using quantum Monte Carlo simulations.
After commenting on hybrid ribbons not aligned along their centers, we recapitulate in Sect.~\ref{sect:conclusion}.


\section{Ribbons of Uniform Width}
\label{sec:ribbons}

\begin{figure}
	\centering
	\includegraphics[width=.48\linewidth]{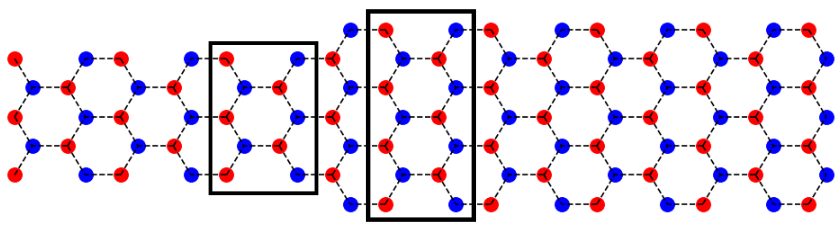}
	\hfill
	\includegraphics[width=.48\linewidth]{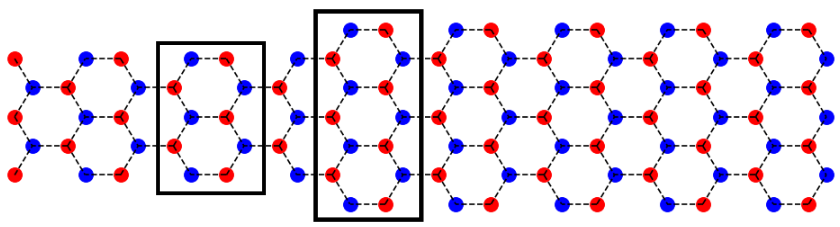}
	\caption{
		A symmetric 5/7 junction made from the intersection of a 5-AGNR and a 7-AGNR.
		The solid rectangles highlight the unit cells of the two individual AGNRs, with two different but equivalent choices shown in the left and right panels.
		The junction resides between the two unit cells shown, respectively.
		Note that the central junction has an additional lattice point residing on the blue sublattice compared to the red sublattice in both cases, as described in the text.
		The central junction and the junction on the edge of the compound unit cell can be thought of as a single unit cell divided in two.
	}\label{fig:7-9-junction}
\end{figure}

Armchair graphene nanoribbons (AGNRs) are carbon nanostructures defined by their edge terminations and can be seen as a portion of an infinite honeycomb lattice with inter-ion spacing $a$.
The ribbons enjoy a translational symmetry along their length which generates a lattice momentum $k$.
The width $N$ of an AGNR is the number of ions along a zigzag path across the ribbon, and a single unit cell consists of two neighboring transverse zigzags.
A ribbon of $m$ unit cells can be compactified with periodic boundary conditions at its ends.
Fig.~\ref{fig:7-9-junction} shows two ribbon segments of widths 5 and 7 joined at a junction.
Clearly both segments as well as the complete hybrid ribbon have a bipartite structure where ions of one triangular sublattice (colored blue)
have neighbors only on the other sublattice (colored red) and vice-versa.

In order to understand how the geometry influences the strength of electronic state localisations, we have to investigate the energy spectra of the different armchair ribbons themselves. 
Of interest will be the state that is closest to zero energy, since this state will govern the long-range correlations.
A gapped system has a finite correlation length while an ungapped system has infinite correlations, cut off in practice by the physical length of the ribbon.

With nearest-neighbor hopping amplitude $\kappa$ these systems are described by the Hamiltonian
\begin{align}
	H = -\kappa \sum_{\langle x, y\rangle} \left(\psi^\dagger_x \psi_y^{} + \psi^\dagger_y \psi_x^{}\right) + \text{interactions}
	\label{eq:hamiltonian}
\end{align}
where $\psi_x$ destroys an electron at site $x$, with $x$ and $y$ are on different sublattices, and we suppress spin labels here and henceforth.
When the interactions are neglected, $H$ is just the tight-binding Hamiltonian used to describe the band structure \cite{band_theory,doi:10.1142/S0217984911025663} and we can find energy eigenstates by diagonalizing the adjacency matrix.

\begin{figure}
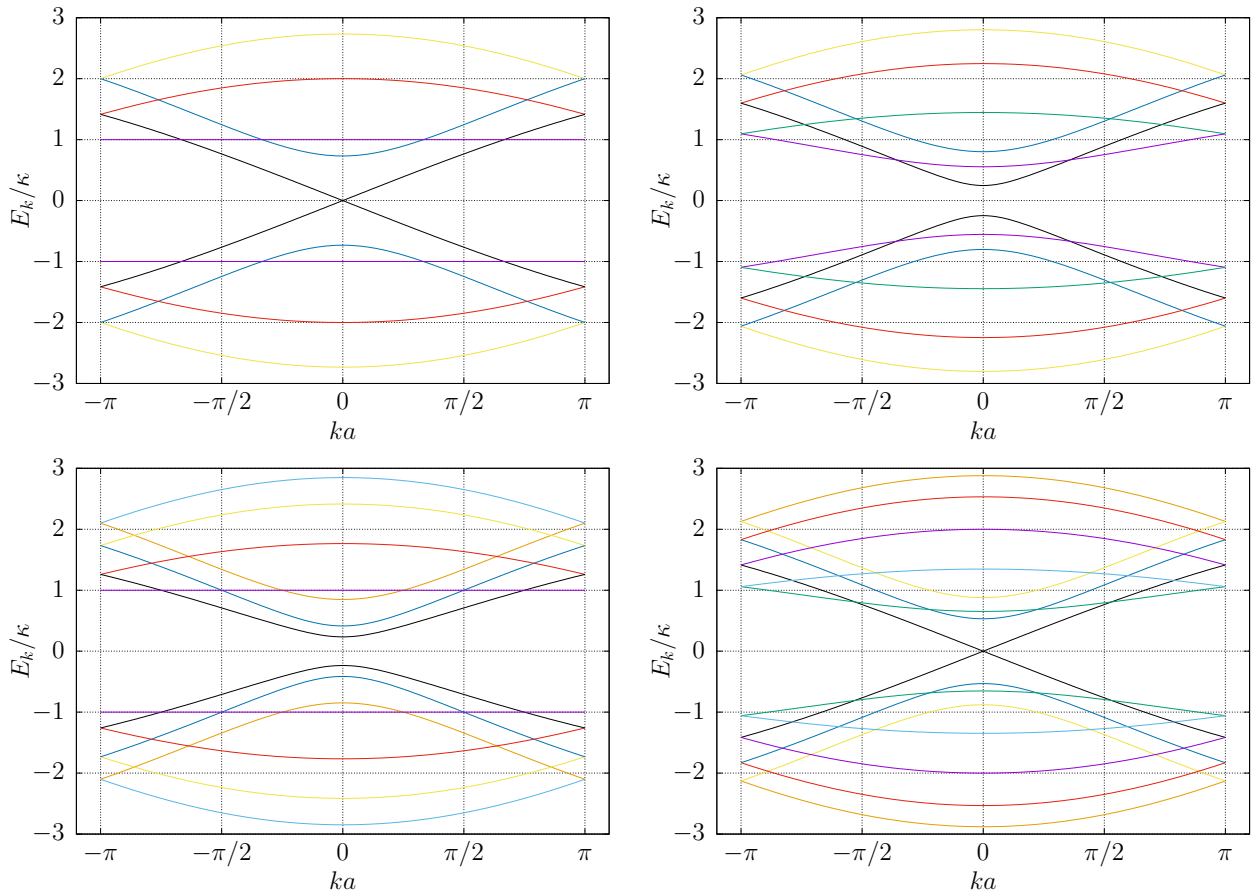

	\centering
	\resizebox{0.95\textwidth}{!}{{\Large\input{data/dispersion-anr5-gnu.tex}\input{data/dispersion-anr6-gnu.tex}}}\\
	\resizebox{0.95\textwidth}{!}{{\Large\input{data/dispersion-anr7-gnu.tex}\input{data/dispersion-anr8-gnu.tex}}}
	\caption{Dispersion relations of infinitely long (i.e.\ $m_N=\infty$) armchair ribbons with widths $N=5,6,7,8$
          (top left to bottom right).}
	\label{fig:dispersions-pure-arm}
\end{figure}

The dispersion relations of armchair ribbons of widths 5 to 8 described by this Hamiltonian are shown in Fig.~\ref{fig:dispersions-pure-arm}.
The armchair ribbons with widths $N=5$ and $N=8$ are gapless while the widths $N=6$ and $N=7$ have finite gaps.
This reflects the well-known fact that armchair ribbons are gapless if and only if their width is
\begin{align}
	N &\equiv 2\pmod{3}. \label{eq:gapless condition}
\end{align}
A general analytic description of the spectrum of these ribbons in the tight-binding model can be found in
Ref.~\cite{Wakabayashi_2010}.
The noninteracting many-body state has all the negative energy states filled.

The authors of~\cite{ribbon_topology} enumerated four distinct types of AGNR edge terminations based on ribbon width
and inversion and mirror symmetries.
They showed that the nanoribbons have an associated binary conserved quantity, the so-called $Z_2$ topological invariant.


\section{Hybrids Ribbons and Junctions}
\label{sec:junctions}

Finite ribbon segments of different width can be joined together to form a \emph{hybrid ribbon}.
The interface of two materials can support surface modes \cite{PhysRevB.95.035421},
in this case modes localized along the hybrid ribbon's length.
We mention two out of the multitude of possible shapes that hybrid ribbons can have: two semi-infinite segments with only a single junction
and repeated segments of alternating widths, with a junction at every width change.
If the alternation is regular the two alternating segments form one compound unit cell which may be repeated
$L$ times along the hybrid ribbon's length; we reuse $m$ to count the number of unit cells in a segment.
The compound unit cell will later be represented by two sites in our effective theory, one site for each junction.

In Ref.~\cite{ribbon_topology}  it was argued that the topology of these systems preserved the localization of states even under the presence of interactions.  
Their perturbative calculations corroborated this claim.  
Consequently in Ref.~\cite{ribbon_junctions} it was shown numerically that this localization persisted in the non-perturbative regime.
In particular,~\cite{ribbon_junctions} investigated the 7/9-hybrid (and the 13/15-hybrid) nanoribbon with non-perturbative
stochastic methods and found that localization indeed persisted in the presence of a Hubbard interaction.
One goal of this present work is to better quantify the nature of these localized states for not only the 7/9 geometry,
but for other hybrid nanoribbon geometries.
As we show in later sections, the dynamics of these low-energy states can be captured in a simple effective 1-D model,
which in turn allows us to make predictions for a broader range of hybrid nanoribbons.  

For simplicity we only consider ribbons segments consisting of a width-$N$ armchair of length $m_N$ and a
width-$N+2$ armchair of length $m_{N+2}$ with odd $N$.
When ribbon segments of different widths are aligned along their centers, as in Fig.~\ref{fig:7-9-junction},
so that the ribbon has a reflection symmetry, the junction has a surplus of a single lattice site, belonging to one of the sublattices
(blue in the center of Fig.~\ref{fig:7-9-junction}, red at the edge).
In this picture it is crucial to tile the hybrid ribbon with unit cells of similar shape in both lattice segments.
The two left-over zigzags on the junctions can be understood as a single unit cell divided.
While in the left panel of Fig.~\ref{fig:7-9-junction} we choose unit cells that are open at top and bottom, we can equivalently choose all
unit cells to be closed as in the right panel.
In the former case the surplus lattice site comes from the junction zigzag of the broad segment while in the latter case the surplus resides
within the narrow segment, but it always belongs to the same sublattice.
This sublattice surplus locally breaks chiral symmetry.
We will find later that hybrid ribbons aligned at an edge do not break chiral symmetry.

\begin{figure}
	\centering
	\includegraphics[width=\textwidth]{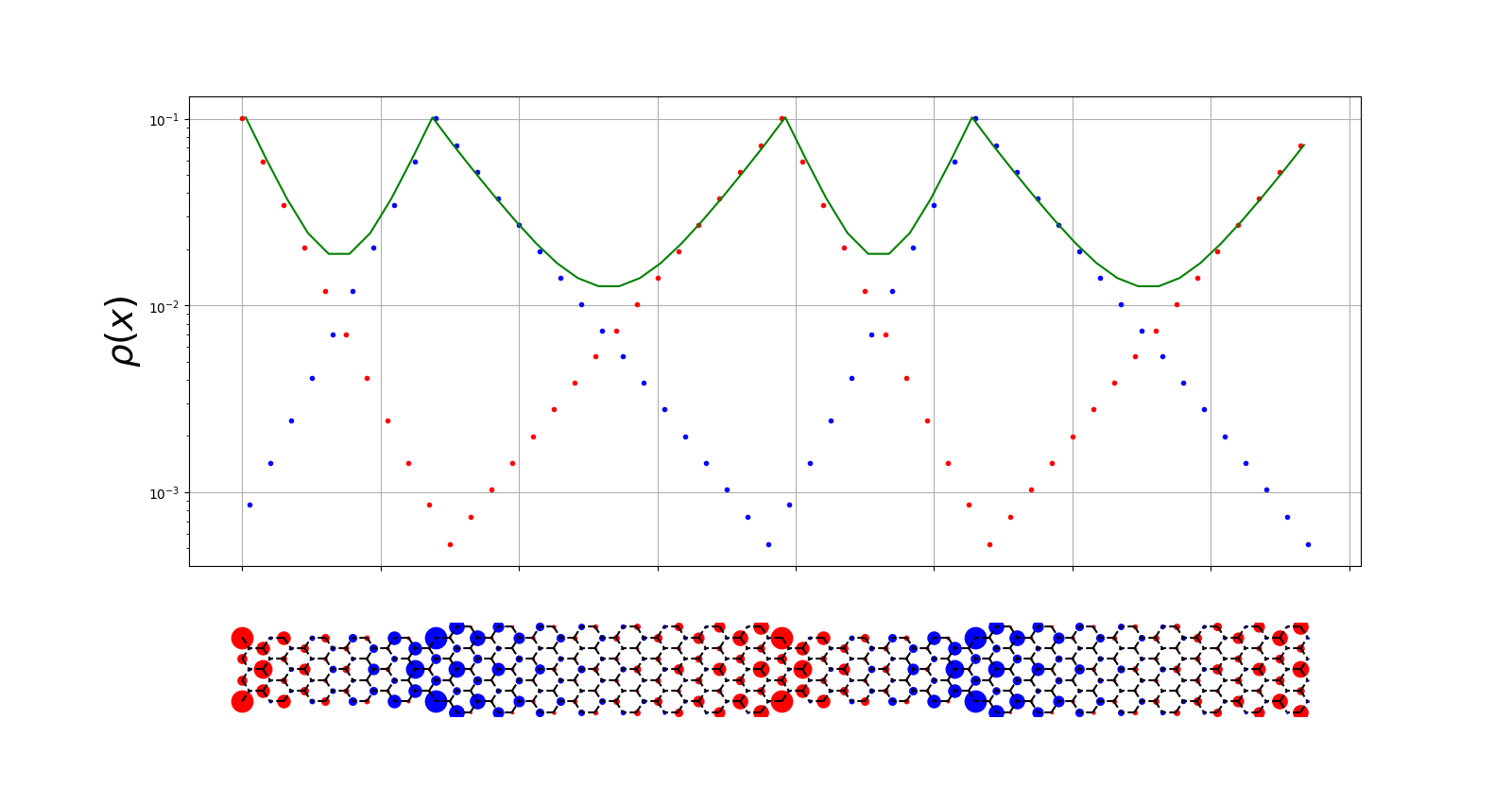}
	\caption{
		Bottom: the \fuj-localised state of a 7/9 hybrid ribbon with $(m_7,m_9)=(5,8)$, shown with $L=2$ two unit cells.
		The circles' radii are proportional to the densities $\rho$ \eqref{eq:rho} and their color indicates the sublattice.
		Top:  We sum $\rho$ along the width of the ribbon and color each point colored according to sublattice. 
		The green line is the sum of both red and blue points along one zigzag cross-section and represents the total occupancy probability (integrated across the ribbon's width) along the ribbon's length.
	}
	\label{fig:7-9-local}
\end{figure}

Fig.~\ref{fig:7-9-local} shows two compound unit cells of an example 7/9-hybrid nanoribbon, where we
see the honeycomb lattice which forms the basis for extended carbon nanostructures.
A ribbon of width $N=7$ has topological invariant $Z_2=0$, while a ribbon of width $N+2=9$ has invariant $Z_2=1$
(more details in Tab.~\ref{tab:topology}); localization is conjectured to occur at the junctions \cite{ribbon_topology}.
This system has been experimentally fabricated~\cite{rizzo18,groening2018}.

Because the geometry controls the gap, a localized state will decay differently on the two sides of the junction.
A localized electron's wavefunction $\phi$ should decay with the dimensionless distance from the junction $\Delta x$.
With large enough length segment length $m$, we expect the asymptotic decay to be governed by the gap or gaplessness
of the infinite ribbon of the same width.%
\footnote{
  Exactly how long each segment needs to be to exhibit such a simple decay is not clear a priori.  While we only
  intend to describe asymptotic behaviour, in practice $m \gtrsim 3$ appears to suffice.
}
In a gapped segment we expect strong localization and exponential decay
\begin{align}
	\phi &\sim  \eto{-\beta \Delta x}\,,\label{eq:strong_localisation}
\end{align}
and in a gapless segment we expect monomial decay
\begin{align}
	\phi &\sim \Delta x^{-\beta}\,,\label{eq:weak_localisation}
\end{align}
and only weak localization.
In both cases $\beta$ is some positive width-dependent parameter independent of segment length $m$ and
the number of compound unit cells $L$.
This dependence on width $N$ has to be determined from fits to solutions of the full problem.

In the bottom panel Fig.~\ref{fig:7-9-local} we show the lowest positive-energy single-electron tight-binding
eigenfunction on a 7/9 hybrid ribbon where the width-7 segments have 5 unit cells and the width-9 segments have 8 unit cells each.
We take the eigenfunction $\phi$ and compute the density normalized per unit cell
\begin{align}
	\rho(x) &= |\phi(x)|^2
	&
	\frac{1}{L} \sum_x \rho(x) &= 1.
	\label{eq:rho}
\end{align}
The radii of the circles are proportional to $\rho$ and colored according to their sublattice.
In the top panel we show the marginal densities $\rho(x)$ summed over the width of the ribbon, again coloring
according to sublattice.
The green line is obtained by adding both the red and blue marginal densities along a transverse zigzag and represents the
total occupancy probability along the ribbon's length.
Both 7- and 9-armchair ribbons are gapped since neither satisfy the gaplessness condition
\eqref{eq:gapless condition}, so correlations decay exponentially on both sides of each junction in Fig.~\ref{fig:7-9-local}.

That the $N=7$ gap is larger than the $N=9$ gap is apparent by the faster decay on the width-7 segments.
We observe that on neighboring junctions the states are not only localized in space but are also concentrated
on one sublattice or the other.
The strong exponential localization allows these states to be clearly delineated.

\begin{figure}
	\centering
	\includegraphics[width=\textwidth]{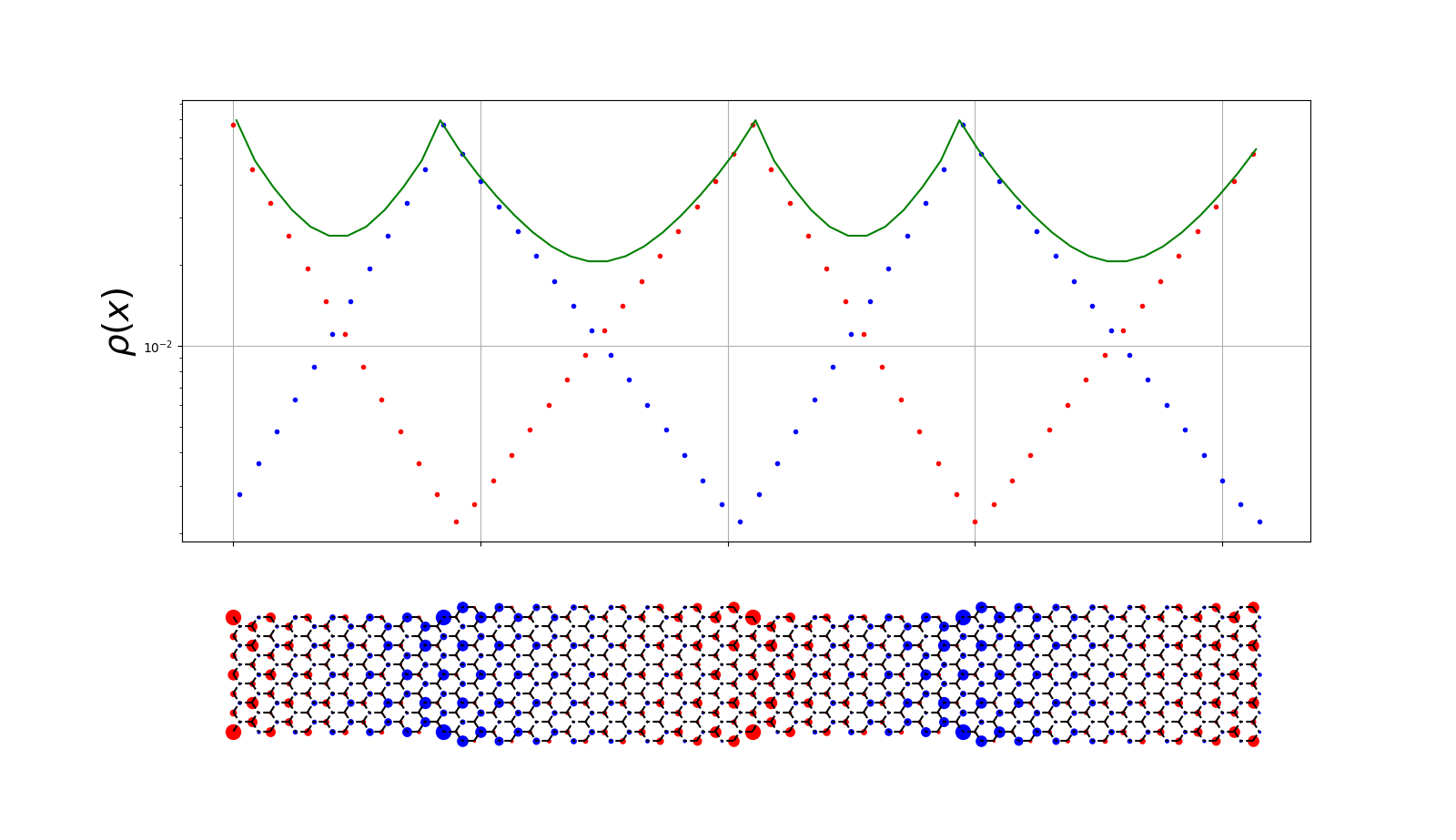}
	\caption{Similar to fig.~\ref{fig:7-9-local} but now the 13/15 hybrid with $(m_{13},m_{15})=(6,8)$.}
	\label{fig:13-15-local}
\end{figure}

We remark that this junction also has changing topology according to Ref.~\cite{ribbon_topology} (see Tab.~\ref{tab:topology})
and their prediction of localisation therefore coincides with ours.
The same occurs for the 13/15 hybrid system, which we show in Fig.~\ref{fig:13-15-local}.
However, we will see that there are counterexamples to the otherwise well-motivated conjecture put
forth in Ref.~\cite{ribbon_topology} that the localizations are driven purely by the topological $Z_2$ boundary.
The model we will develop in Sec.~\ref{sec:theory} is generally applicable and reliably quantifies
localizations even in the cases that evade the topological argument.

\begin{figure}
	\centering
	\includegraphics[width=\textwidth]{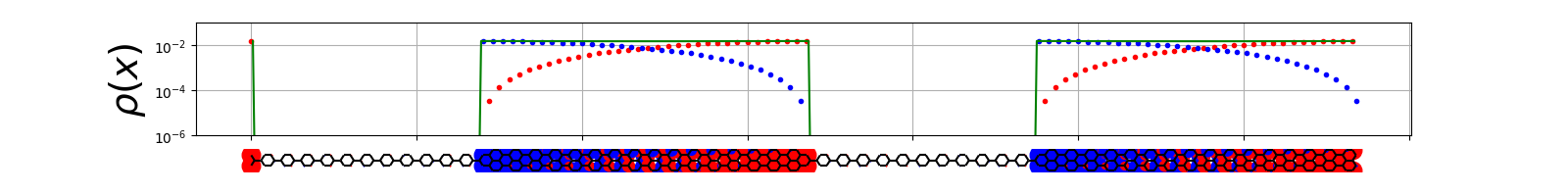}
	\includegraphics[width=\textwidth]{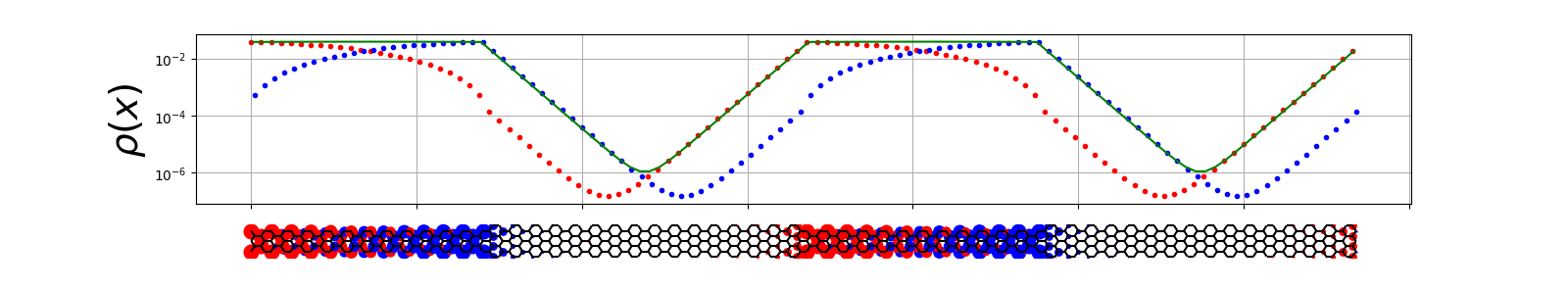}
	\includegraphics[width=\textwidth]{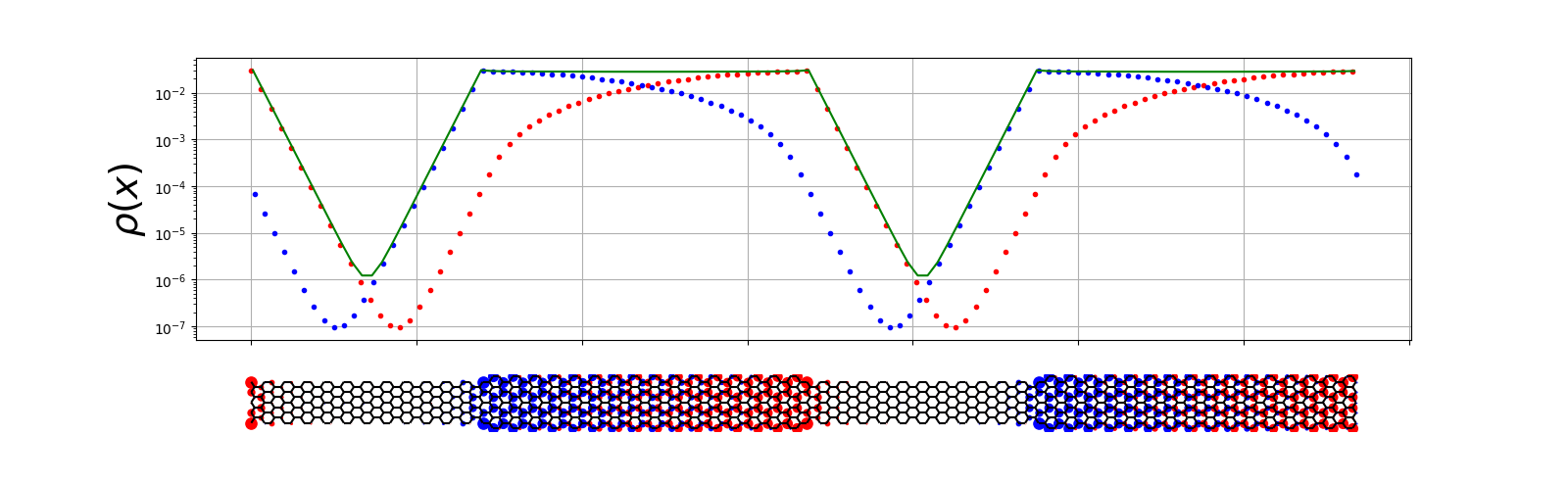}
	\caption{
		Lowest energy state densities of a 3/5 (top), a 5/7 (middle) and a 9/11 hybrid (bottom), all with segment lengths $(m_N,m_{N+2})=(12,16)$.
		These examples do not feature the two-sided exponential \fuj localisation on the junctions since width-5 and width-11 armchair ribbons have long range correlations.
		However, the states are trapped within those gapless segments, showing \kil localisation.
	}
	\label{fig:kilimanjaro}
\end{figure}

Fig.~\ref{fig:kilimanjaro} shows the low-energy states from 3/5, 5/7, and 9/11 hybrid ribbons.
Each of these examples has a gapless segment, since $5\equiv11\equiv2\pmod{3}$ satisfying the gaplessness
condition \eqref{eq:gapless condition}, and on the gapless segment no sharp localization on the junction occurs.
Instead, on the scale shown the eigenstate looks essentially constant on the gapless segments.

We distinguish these `\kil-localized' states with a large plateau from the sharply-peaked `\fuj-localized'
states that have exponential decay on both sides of a junction.\footnote{Mount \kil in Tanzania has an extended high plateau, while the Japanese mount \fuj features a sharp peak. The resemblances to the respective localisations inspired the naming scheme.}
We remark that the cumulated occupancy density shown in green is not exactly constant in the plateau region.
Instead, the density increases towards the center.
In fact, if the gapless segment is very short, the localization can be very sharp, not unlike \fuj localization.
But, the state can also be meaningfully spread over vast regions if the gapless segment is long enough.

Focusing on the 5/7 hybrid, as we make $m_7$ larger the low-energy state remains confined to the width-5 segments.
If we take $m_7 \gg m_5$, we can effectively localize the density into an arbitrarily small space compared to the total
length of the ribbon.
Unlike the \fuj localization, in this limit there is no sharp splitting between the two sublattices.
The localization in the gapless segment are only polynomial in nature and states localized to the two sublattices at
either end of the gapless segment have a large overlap.

The 5/7 example, in particular, contradicts the claim in Ref.~\cite{ribbon_topology} that a change in topology implies
a \fuj localization.
However, we find that the reverse implication---localization requires a change in topology---is consistent with the
examples we have examined and the effective theory we present in Sec.~\ref{sec:theory}.

The findings of Ref.~\cite{ribbon_topology} are based on hybrid ribbons with a single junction connected by semi-infinite ends, whereas our investigations here involve hybrid ribbons with periodic boundary conditions, which essentially models an infinite number of junctions.
A natural question is whether this difference accounts for the discrepancy between our findings.
With our numerical techniques it is not possible to model infinite ribbons.
However, instead of periodic boundary conditions at the ends, we can use open boundary conditions and investigate the nature of the localization as we extend the length of each semi-ribbon.
\begin{figure}
	\centering
	\includegraphics[width=\textwidth]{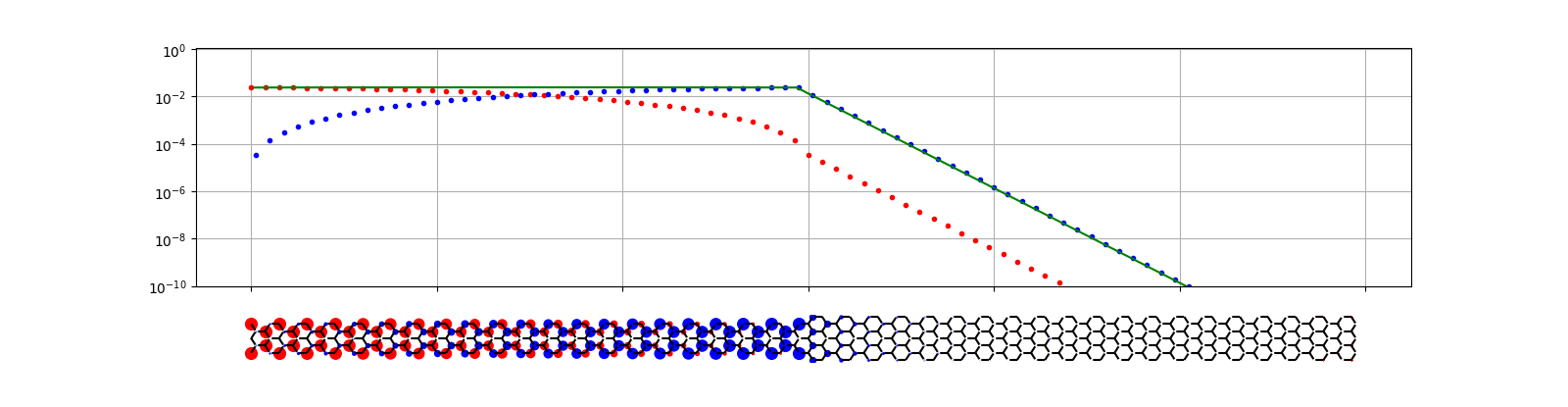}
	\includegraphics[width=\textwidth]{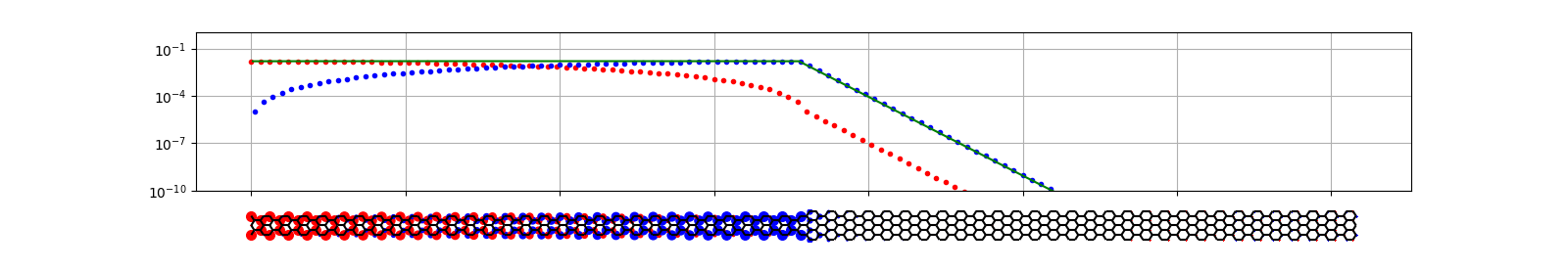}
	\includegraphics[width=\textwidth]{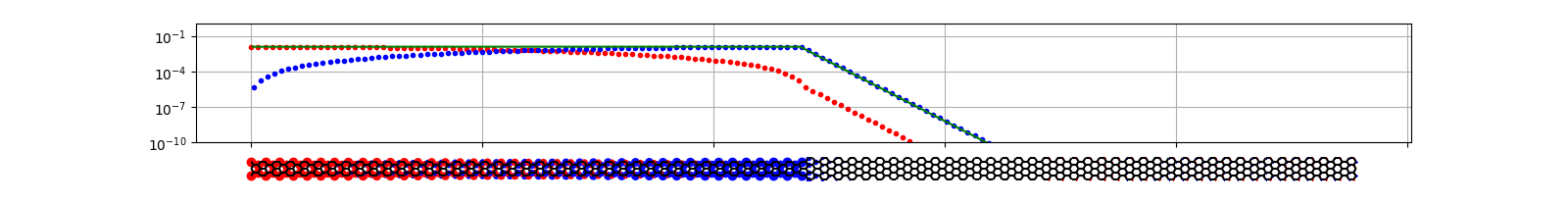}
	\caption{
		Lowest non-zero energy state densities of the 5/7 ribbon with open boundary conditions. The top panel has $(m_N,m_{N+2})=(20,20)$, middle $(m_N,m_{N+2})=(30,30)$, and bottom $(m_N,m_{N+2})=(40,40)$.
		The states are again trapped within those gapless segments and demonstrate \kil localisation.
	}
	\label{fig:open}
\end{figure}
We show the length-normalized densities for the lowest non-zero energy state\footnote{
	With open boundary conditions we always find two degenerate zero-energy states that correspond to perfect localisations on the extreme ends of the system.  These states play no role in the limit of semi-infinite ends, as their localisations are pushed to infinity.  The relevant states are the lowest non-zero energy states, which we show in Fig.~\ref{fig:open}
} for increasingly long 5/7 hybrid ribbons with open boundaries in Fig.~\ref{fig:open}.
The \kil localisation is prominent and remains so as the ribbons' respective lengths increase.  
We therefore surmise that this type of localisation persists in the limit of semi-infinite ends. This is perfectly in line with the expectations in our ET framework and cannot be reconciled with the predictions in Ref.~\cite{ribbon_topology}.



\begin{table}[h]
\caption{Topological invariant~\cite{ribbon_topology} (Tab.~I therein) for the narrower and broader parts
          of different junctions respectively. In a hybrid ribbon with a symmetric junction (\cref{fig:7-9-junction,fig:7-9-local,fig:13-15-local,fig:kilimanjaro}),
          the $Z_2'$ invariant describes the topology in the narrow and $Z_2$ the broader segment.
          In bottom aligned junctions (\cref{fig:9-11-shift}) both parts are described by the $Z_2'$ invariant.
          Ref.~\cite{ribbon_topology} predicts localisations for junctions with
          changing topology.}\label{tab:topology}
          \centering
	\begin{tabular}{|l|*{10}{S}|}
		\hline
		$N$ & 3 & 5 & 7 & 9 & 11 & 13 & 15 & 17 & 19 & 21 \\\hline
		$N+2$ & 5 & 7 & 9 & 11 & 13 & 15 & 17 & 19 & 21 & 23 \\\hline
		$Z_{2}'{(N)}$ & 1 & 0 & 0 & 0 & 1 & 1 & 1 & 0 & 0 & 0 \\\hline
		$Z_{2}{(N+2)}$ & 1 & 1 & 1 & 0 & 0 & 0 & 1 & 1 & 1 & 0 \\\hline
	\end{tabular}
	\end{table}


\section{Effective 1-D Tight-binding model}
\label{sec:theory}

\subsection{Formulation}

An electron localized on a junction is smeared out over many sites of one sublattice near by.
We observe in Figs.~\ref{fig:7-9-local}, \ref{fig:13-15-local}, and \ref{fig:kilimanjaro} that at a junction the wavefunction is concentrated on the sublattice with a surplus site.
This sublattice symmetry breaking and wavefunction concentration allows us to treat the $2L$ junctions from $L$ compound unit cells as the sites of our model.
Because the junctions alternate between having a surplus of one of the honeycomb sublattices (and the corresponding
wavefunction  concentration), we arrive at a length $L$ bipartite lattice with a two-site basis.
The two effective sites can be thought of as the local surplus of one or the other sublattice.
Electrons hop between these effective sites via some hopping amplitude controlled by the width and length of the segment
connecting them; a segment of width $N$ and length $m_N$ lets electrons tunnel with an amplitude controlled by the
wavefunction overlap.
If two junctions are separated by a strongly-localizing segment \eqref{eq:strong_localisation}
of length $m$ the wavefunction overlap and thus the tunnelling amplitude $t$ will be exponentially small,
\begin{align}
	t &\sim \eto{-\beta m}\,,
	\label{eq:exponential overlap}
\end{align}
while two junctions separated by a weakly-localizing segment \eqref{eq:weak_localisation} will have polynomial overlap and tunnelling amplitude
\begin{align}
	t &\sim m^{-\beta}\,,
	\label{eq:polynomial overlap}
\end{align}
redefining the dimensionless $\beta$.

An effective 1-D tight binding Hamiltonian that describes a hybrid ribbon of alternating widths $N$ and $N+2$ is
\begin{align}
	H_\text{1D}
	&=
	-\sum_{x=0}^{L-1} \left(
		t_N c^\dagger_{2x}c^\pdagger_{2x+1} + t_{N+2} c^\dagger_{2x+1}c^\pdagger_{2x+2} + {\rm h.c.}
	\right)\,,
	\label{eq:1d_Hamiltonian}
\end{align}
where $c_x$ destroys a fermion at effective site $x$, and $t_{N}$ is the tunnelling (or hopping) amplitude across a ribbon segment of width $N$.
It can be block-diagonalised by a Fourier transformation yielding
\begin{align}\label{eqn:1D H0}
  H_\text{1D} &= -\sum_k c^\dagger_k \matr{cc}{0&t_N\eto{\im k}+t_{N+2}\eto{-\im k}\\t_N\eto{-\im k}+t_{N+2}
    \eto{\im k}&0} c^\pdagger_k\,,
\end{align}
where the dimensionless momentum $k$ is in terms of the inverse lattice spacing and the creation and annihilation
operators in momentum space are two-dimensional vectors,
\begin{equation}
c_k=
	\begin{pmatrix}
	c_{k,A}\\
	c_{k,B}
	\end{pmatrix}
\end{equation}
and the $A$ and $B$ indices indicate the two sublattices or equivalently the two junctions.

After diagonalising the blocks we obtain the dispersion relation
\begin{align}
	E(k) &= \pm\sqrt{t_N^2+t_{N+2}^2+2t_Nt_{N+2}\cos 2k}
\end{align}
for momenta in the reduced first Brillouin zone $k\in[0,\pi)$  and the energy gap
\begin{align}
	\Delta &\equiv 2|E(\pi/2)| =2\sqrt{t_N^2+t_{N+2}^2-2t_Nt_{N+2}}
	= 2\left|t_N-t_{N+2}\right|\label{eq:junction_gap}
\end{align}
between lowest positive and highest negative energies which will become very important in the following considerations.
Note that a hybrid ribbon with small $t_N$ and $t_{N+2}$ necessarily has a small gap.
However, a small $t_{N}$ is a consequence of a large pure-armchair gap since in this case it less likely to hop between junctions.
This effective theory predicts that joining two strongly gapped ribbons leads to a very small overall gap. 

Sharpening the scaling of the overlaps \eqref{eq:exponential overlap} and \eqref{eq:polynomial overlap} into
quantitative predictions, the effective hopping amplitudes are
\begin{align}
	t_N(m) &= \begin{cases}
		\kappa\, \alpha\, m^{-\beta} \;\quad \text{with } \beta \sim 1\,, & \text{if } N = 2\pmod 3\,,\\
		\kappa\, \alpha\, \eto{-\beta m} \quad \text{with } \beta \sim \Delta_N\,, & \text{otherwise,}
	\end{cases}
	\label{eq:effective_hopping}
\end{align}
with $\alpha$ another (apriori unknown) positive dimensionless parameter that can only depend on $N$, not
on $m$ and $L$ and the honeycomb $\kappa$ \eqref{eq:hamiltonian} appears for dimensional reasons.
In the first case $\beta$ is expected to be related to critical behaviour and cannot be predicted from first principles.
In contrast, the exponential decay is governed by the magnitude of the pure $N$-armchair ribbon gap $\Delta_N$ up to
small corrections.
We will use this ansatz to fit the low-energy constants (LECs) $\alpha$ and $\beta$ for different values of $N$.

Concisely, the effective treatment predicts that an $N$/$N+2$ hybrid ribbon of two armchair nanoribbons has
\fuj-localised states with close to zero energy if and only if the junction is center-aligned and $N\pmod3=1$ so
that neither width fulfils the gaplessness condition \eqref{eq:gapless condition}.

\subsection{Determination of the Low-Energy Constants}
	
\begin{figure}
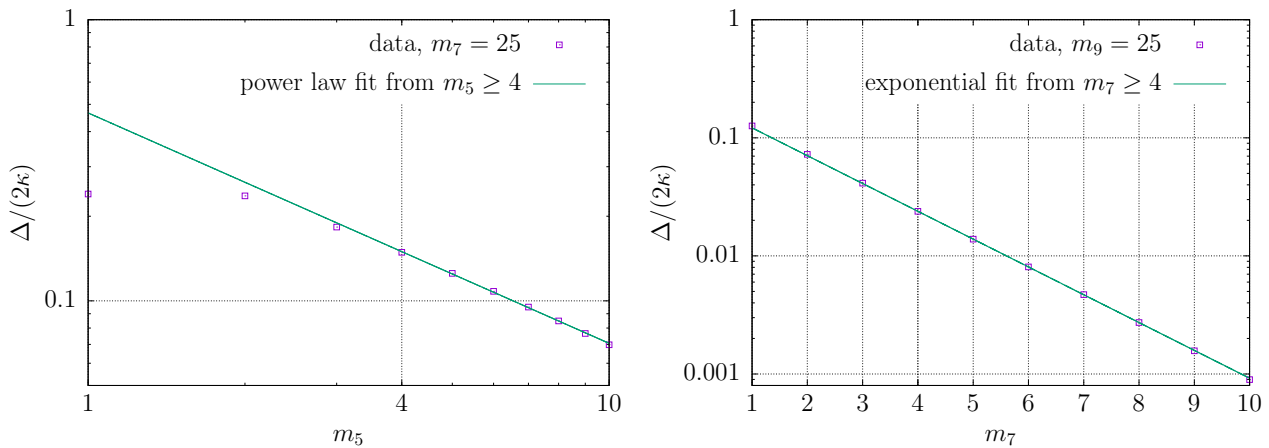

	\resizebox{0.95\textwidth}{!}{{\Large\input{data/ribbons_5-7}\input{data/ribbons_7-9}}}
	\caption{Gaps of a 5/7-hybrid (left) and a 7/9-hybrid (right) used to fit the LECs \eqref{eq:effective_hopping} for $N=5$ and $N=7$,
          respectively. When fitting a power law as a function of the length $m_5$,
          the length $m_7=25$ has been kept fixed and similarly for the exponential fit to $m_7$ we fixed $m_9=25$.}
        \label{fig:fit_5-7-9-junction}
\end{figure}

We now have all ingredients to fix the low-energy constants \eqref{eq:effective_hopping} of our 1-D effective
theory \eqref{eq:1d_Hamiltonian}.
By considering a particular $N/N+2$ hybrid ribbon, we calculate the gap $\Delta$ (defined as twice the lowest positive
single-particle energy) of the hybrid system for different ribbon lengths $m_N$ and $m_{N+2}$.
For the sake of simplicity we choose one of the lengths very large, say $m_{N+2} \gg m_N$, so that the $N+2$-width
ribbon segment is long enough to be compatible with the thermodynamic limit.
Then the effects of this ribbon segment are negligible and the junction gap~\eqref{eq:junction_gap} reduces to
$\Delta=2t_N$.
We fit our results for $t_N(m_N)$ to the form of the effective hopping~\eqref{eq:effective_hopping}, fixing the
parameters $\alpha_N$ and $\beta_N$.
Two representative fits are shown in Fig.~\ref{fig:fit_5-7-9-junction}, with a power law fit in the left panel
and an exponential fit on the right.

We summarise the results of the fitted low-energy constants in table~\ref{tab:LECs} for select values of $N$.
Within either class, power law or exponential, we observe the trend that both LECs $\alpha$ and $\beta$
decrease with growing $N$. While we do not have a direct physical interpretation for the proportionality
constant $\alpha$, it is clear that $\beta$ has to follow this trend because the asymptotic $N\rightarrow\infty$
case of graphene is gapless. In particular, the exponential case features decay coefficients $\beta$ similar
to the pure armchair ribbon gap $\Delta_N$ as expected. 

Note how the 7/9-junction is special in the sense that it is the smallest ribbon size with strong localisation for both widths.
No \fuj\ localisation is possible in narrower center-aligned ribbons.
We also remark that the 3-armchair ribbon features such a strong exponential decay that it is virtually instant and
(at least within double floating precision) $t_3(m) = 0$ for $m>0$. Localised states do not penetrate into the 3-armchair at all.

\begin{table}[h]
        \caption{Fitted low-energy constants (LECs) $\alpha$, $\beta$ from eq.~\eqref{eq:effective_hopping} following the exponential (exp)
          or power (pow) laws depending on the width $N$ of the armchair ribbon. $\Delta_N$
          is the corresponding energy gap of the ribbon without junction. For $N=3$ we have $\beta=\infty$ and there is no value for
          $\alpha$ since the wavefunction is exactly confined to the junction (see Fig.~\ref{fig:kilimanjaro}).\label{tab:LECs}}
	\centering
	\begin{tabular}{|c|*{10}{S[round-mode = places,round-precision = 2,round-pad = false]}|}
		\hline
		$N$ & 3 & 5 & 7 & 9 & 11 & 13& 15& 17& 19& 21\\
		\hline
		Decay & \text{exp} & \text{pow} & \text{exp} & \text{exp} & \text{pow} & \text{exp} & \text{exp} & \text{pow} & \text{exp} & \text{exp}\\
		\hline
		$\alpha$ & \text{-} & 0.565507 &0.208839 &0.223853 & 0.430726 &0.129654 &0.147543 & 0.318031 &0.0936847 & 0.107584 \\
		\hline
		$\beta$ & $\infty$ &0.885373 &0.542315 &0.339993 & 0.794006 &0.302953 &0.232692 &0.688998 &0.219054 &0.179434 \\
		\hline
		$\Delta_N$ & 0.828427& 0 & 0.469266 & 0.351141 & 0 & 0.264465 & 0.222281 & 0 & 0.184038 & 0.162563\\
		\hline
	\end{tabular}

\end{table}

\subsection{Application of our effective theory}

Despite the simplicity of our effective theory, we can already use it to make predictions in cases where the original
system is more difficult to simulate.
We can apply our ET, for example, to predict the respective lengths at which the gap of a hybrid nanoribbon
(almost) vanishes. 
As an example we return to our prototypical 7/9 hybrid system, but with the desire to pick segment lengths so that
the system is as close as possible to gapless.

\begin{figure}
	\input{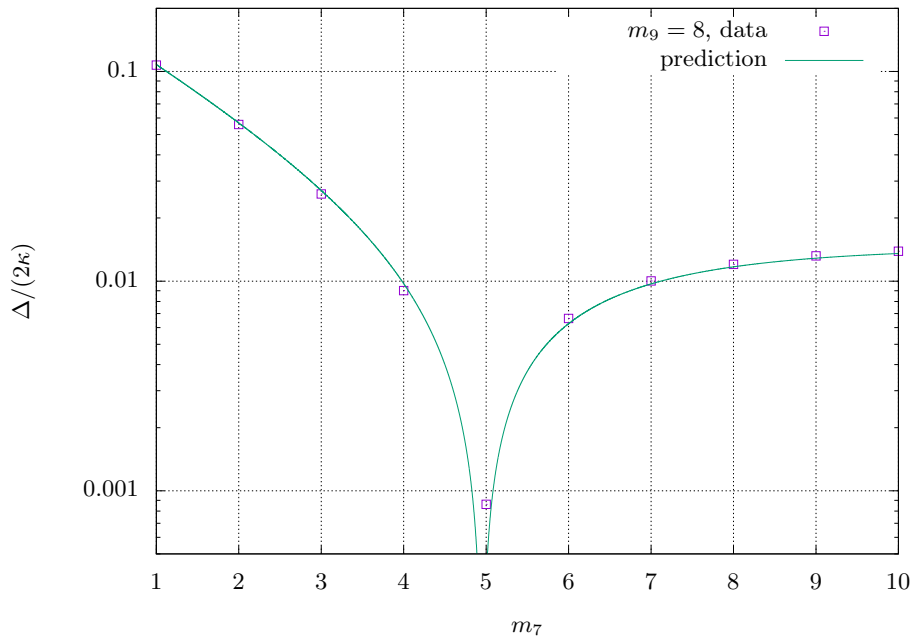}
	\caption{
	  Gap of a 7/9-hybrid ribbon given by the prediction~\eqref{eq:predict_min_gap} and direct diagonalization
          of the underlying tight-binding Hamiltonian shown in~\eqref{eq:hamiltonian} (without interactions).
	}
	\label{fig:min-gap}
\end{figure}

To minimize the gap \eqref{eq:junction_gap} our ET provides the condition
\begin{align}
	t_7 &\overset{!}{=} t_9
	&&\Rightarrow
	&	
	\alpha_7\eto{-\beta_7 m_7} 
	&\overset{!}{=}
	\alpha_9\eto{-\beta_9 m_9}
	&&\Rightarrow
	&
	m_7
	&=
	\frac{\beta_9}{\beta_7}m_9 + \ln\frac{\alpha_7}{\alpha_9}
	\label{eq:predict_min_gap}
\end{align}
has to hold as best possible for integers $m_7$ and $m_9$.
Using the parameters given in Table~\ref{tab:LECs} we find that $\left(m_7,m_9\right) = \left(5,8\right)$  is a
good tuple that nearly satisfies this constraint.
This prediction is confirmed in Fig.~\ref{fig:min-gap}, which shows the hybrid ribbon's gap as a function of
the width-7 segments' length, holding the width-9 segments at $m_9=8$.
The next three smallest tuples that our theory predicts for this system are $(22,35)$, $(39,62)$, and $(56,89)$.
For the 13/15 hybrid system our effective theory predicts the following four smallest tuples giving a near
zero gap: $(m_{13},m_{15})$=$(6,8)$, $(29,38)$, $(52,68)$, and $(75,98)$.  

Note that in both these systems, both ribbon widths are gapped and the localization is \fuj.
For systems where one width is gapped and the the other is not, our theory predicts that such systems cannot
support a (near) zero gap without weakly-localising segments many orders of magnitude longer than the strongly-localising
segments. This is consistent with all our simulations to date.

\subsection{Incorporating Interactions}
\label{sect:interactions}
	
So far we have focused on noninteracting tight-binding dynamics, both within the hybrid nanoribbon and its effective
1-D description.
Including interactions, for example by adding an onsite Hubbard interaction $U$ that couples the spin-up $\uparrow$ and
spin-down $\downarrow$ electrons
\begin{equation}
	\label{eq:hubbard}
	H_{\text{Hubbard}} = U\sum_{x}\left(\psi^\dagger_{x,\uparrow}\psi^{}_{x,\uparrow}-\frac{1}{2}\right)
        \left(\psi^\dagger_{x,\downarrow}\psi^{}_{x,\downarrow}-\frac{1}{2}\right)
\end{equation}
to the underlying tight-binding Hamiltonian \eqref{eq:hamiltonian}, precludes simple diagnolization.

\begin{figure}
	\includegraphics[width=.8\textwidth]{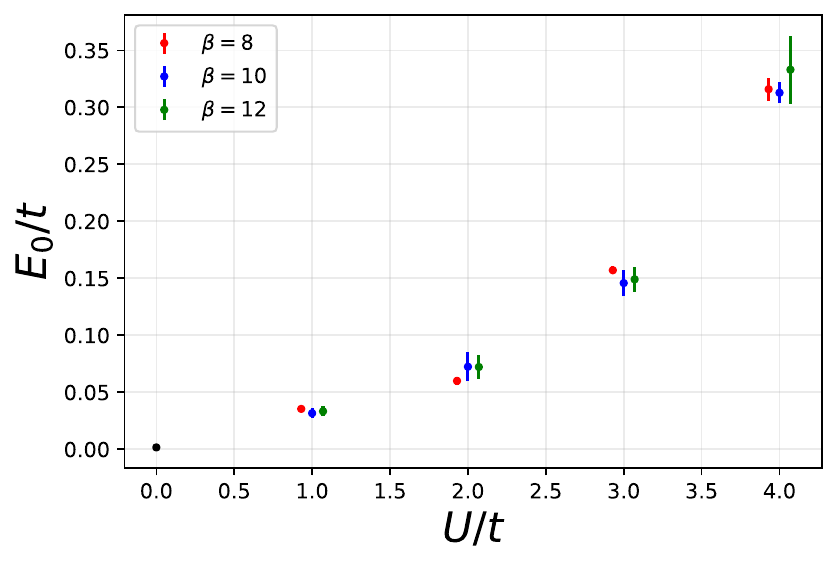}
	\caption{
		Interacting energy $E_0$, depicted as points with error bars, of the lowest state as a function of onsite Hubbard interaction $U$ obtained from QMC calculations in Ref.~\cite{ribbon_junctions} for the $7/9$ system with $(m_7,m_9)=(3,5)$.
		The $7/9$ simulations were performed with three different values of inverse temperature $\beta$, where $\beta=8$ (12) results are slightly shifted to the left (right) to help visually differentiate the points.
		The black point corresponds to the non-interacting result.
	}
	\label{fig:E vs U}
\end{figure}

Ref.~\cite{ribbon_junctions} showed that the localization was robust against the influence of the Hubbard
interaction \eqref{eq:hubbard} via stochastic Monte Carlo methods and that there is a nearly quadratic dependence
of the gap on $U$. 
Fig.~\ref{fig:E vs U} shows this dependence for the example of the 7/9 system.

Our 1-D effective model \eqref{eq:1d_Hamiltonian} can easily incorporate these results by including
\begin{equation}
	m_s\left(c^\dagger_{2x}c^\pdagger_{2x} - c^\dagger_{2x+1}c^\pdagger_{2x+1}\right)\ ,
\end{equation}
where the effective staggered mass $m_s$ is a LEC and fit to reproduce the quadratic dependence.
The momentum-space formulation \eqref{eqn:1D H0} becomes
	\begin{align}
	  H_\text{1D} &= -\sum_k c^\dagger_k \matr{cc}{m_s&t_N\eto{\im k}+t_{N+2}\eto{-\im k}\\
            t_N\eto{-\im k}+t_{N+2}\eto{\im k}&-m_s} c^\pdagger_k\,,
	\end{align}
which can be easily diagonalized, giving
\begin{align}
	E(k) &= \pm\sqrt{t_N^2+t_{N+2}^2+2t_Nt_{N+2}\cos 2k+m_s^2}\ .
\end{align}
and a gap
\begin{equation}\label{eqn:ET gap}
\Delta = 2|E(\pi/2)|=2\sqrt{(t_N-t_{N+2})^2+m_s^2}\ .
\end{equation}
The presence of this staggered mass does not change the scaling behavior of the hopping terms~\eqref{eq:effective_hopping}
and therefore does not affect the nature of the localization.
For a given $U$ simulated with a particular tuple $(m_N,m_{N+2})$, the parameter $m_s$ can be tuned so that
our ET matches the energy of the underlying theory, like that shown in Fig.~\ref{fig:E vs U}. 
Once tuned, we can then make predictions for the size of the gap for hybrid ribbons with segments of the same
widths but with different lengths.

The tuple that minimizes the gap will be the one that corresponds to $|t_N-t_{N+2}|\sim 0$.
Since the staggered mass preserves the scaling behavior of the hopping terms, the predicted tuples that minimize
the gap in the previous section when $m_s=0$ will also minimize the gap for $m_s\ne 0$.
However, in this case the minimum gap becomes $\Delta\sim 2m_s$.

\begin{figure}
\includegraphics[width=.8\textwidth,]{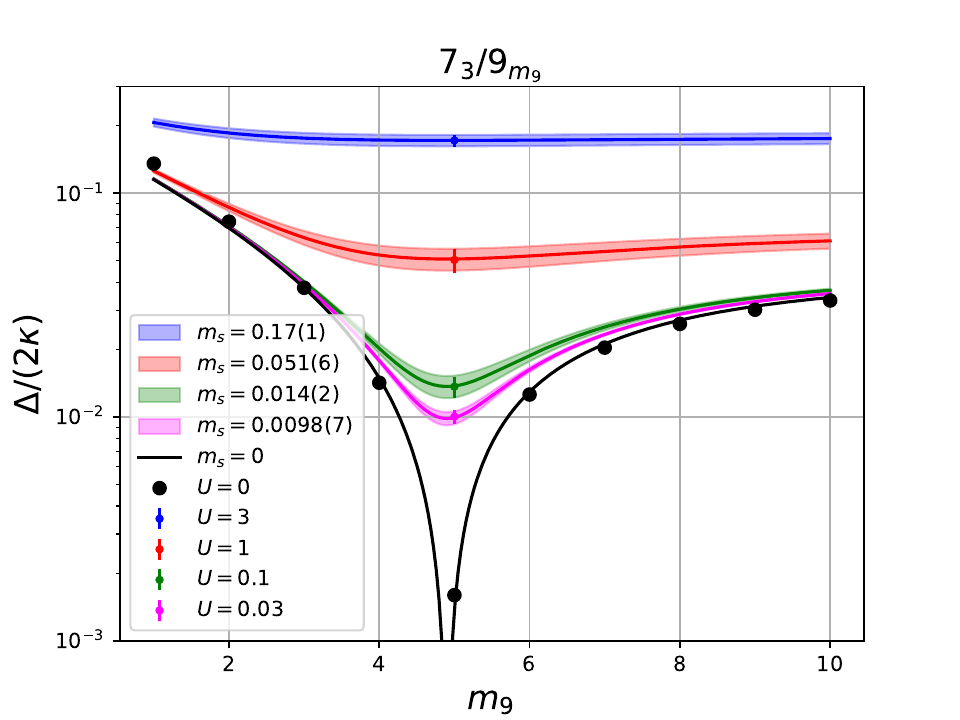}
\caption{Extracting $m_s$ from the underlying theory.  Here stochastic simulations of the full $7/9$ hybrid system with tuple $(m_7,m_9)=(3,5)$, $L=1$ and inverse temperature $\beta=8$ were performed at different values of $U$ as labelled in the figure and shown as points with error bars.  The value of $m_s$ was fitted to each of these points, and the resulting prediction of the gap provided by our ET (eq.~\ref{eqn:ET gap}) for other tuples where $m_7=3$ and $m_9\in[1,10]$ is plotted.  The black points are the non-interacting results. \label{fig:79 fit interactions}}
\end{figure}

\begin{figure}
\includegraphics[width=.8\textwidth]{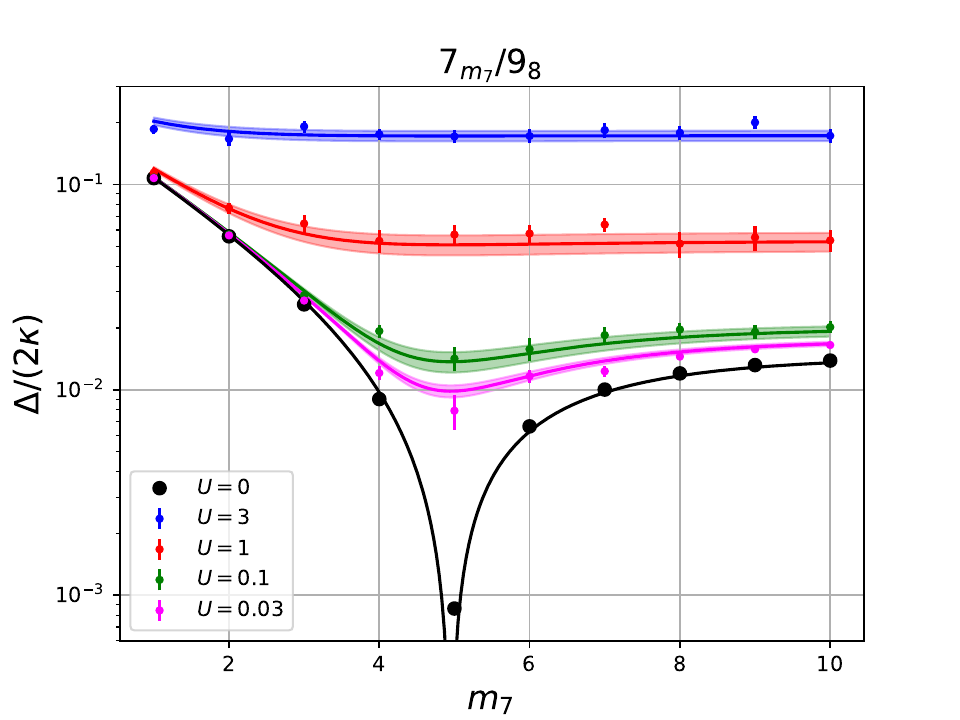}
\caption{Comparing our ET prediction with the underlying theory.  Using the values of $m_s$ extracted in Fig.~\ref{fig:79 fit interactions}, we plot our ET prediction of the gap, shown as bands, for $7/9$ geometries where $m_9=8$ and $m_7\in[1,10]$.  Superimposed on these bands are the gaps obtained from stochastic simulations of the underlying theory of these systems.  
 \label{fig:79 w/ interactions}}
\end{figure}

As an example of how we can extract $m_s$, we perform stochastic simulations of the underlying Hubbard theory on the full $7/9$ hybrid ribbon with tuplet $(m_7,m_9)=(3,5)$.
The details of our Quantum Monte Carlo (QMC) simulations are described in~\cite{ribbon_junctions}.
In short, we sample the electron configurations from their quantum mechanical probability distribution using a Markov chain with global updates. In the limit of high statistics these simulations become exact. Given limited computational resources, we arrive at a distribution of values around the true result and we depict the standard error of this distribution as error bars in \Cref{fig:E vs U,fig:79 fit interactions,fig:79 w/ interactions}.

The results of the gap for different values of Hubbard coupling $U$ are shown as points with errorbars in Fig.~\ref{fig:79 fit interactions}.
We then fit our ET to these results, thereby extracting $m_s$ with the values shown in Fig.~\ref{fig:79 fit interactions}. 
With $m_s$ in hand, we can predict the value of the gap for other combinations of segment lengths, shown by bands in the same figure.

To demonstrate the efficacy of our ET, we use these same values of $m_s$ to plot our predicted gaps for completely different $7/9$ geometries, with $m_9=8$, in Fig.~\ref{fig:79 w/ interactions}.  
Every band in Fig.~\ref{fig:79 w/ interactions} is a prediction given the low energy constants $\alpha$ and $\beta$ from the noninteracting case and the effective staggered mass $m_s$ for that Hubbard coupling.
In particular, the $(m_7, m_9)=(3,5)$ hybrid geometry used to extract $m_s$ does not appear in Fig.~\ref{fig:79 w/ interactions} at all.
We then perform stochastic simulations of the underlying theory of these systems and plot their resulting gaps, shown as data points with errorbars.
We find good agreement between our simulations and ET.  
We thus surmise that our ET with a staggered mass captures both the dynamics and interactions of the lowest energy spectrum of the hybrid nanoribbons.

More quantitative descriptions of interacting hybrid nanoribbons, potentially going beyond Hubbard interactions, are possible within our formalism.
For example, the inclusion of off-diagonal superconducting pairing terms, i.e.\ $c_kc_k$ and $c^\dag_kc^\dag_k$, may be done with the aid of a Bogoliubov transformation~\cite{Bogoljubov}.
One could alter the dynamics of the system by including next-to-nearest neighbor hoppings, or extend the interaction by considering onsite plus nearest neighbor couplings (i.e.\ extended Hubbard).
Such possibilities are the subject of future investigations.

\subsection{Misaligned Hybrid Ribbons}

\begin{figure}
	\centering
	\includegraphics[width=\textwidth]{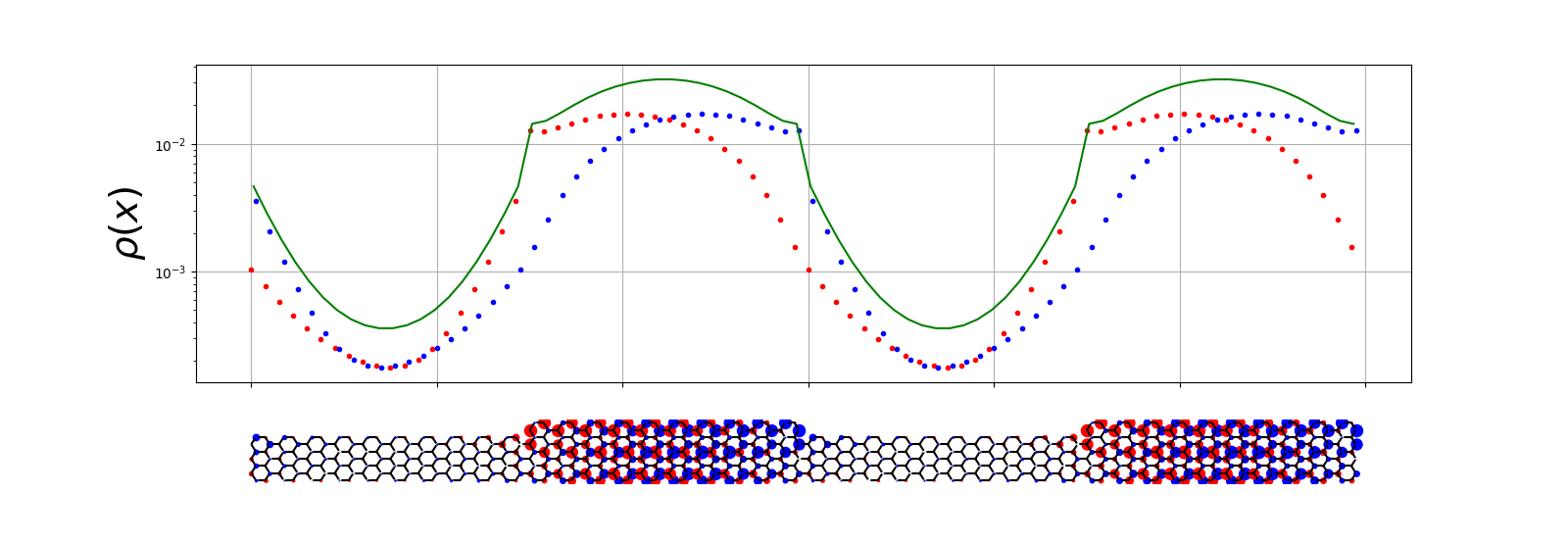}
	\includegraphics[width=\textwidth]{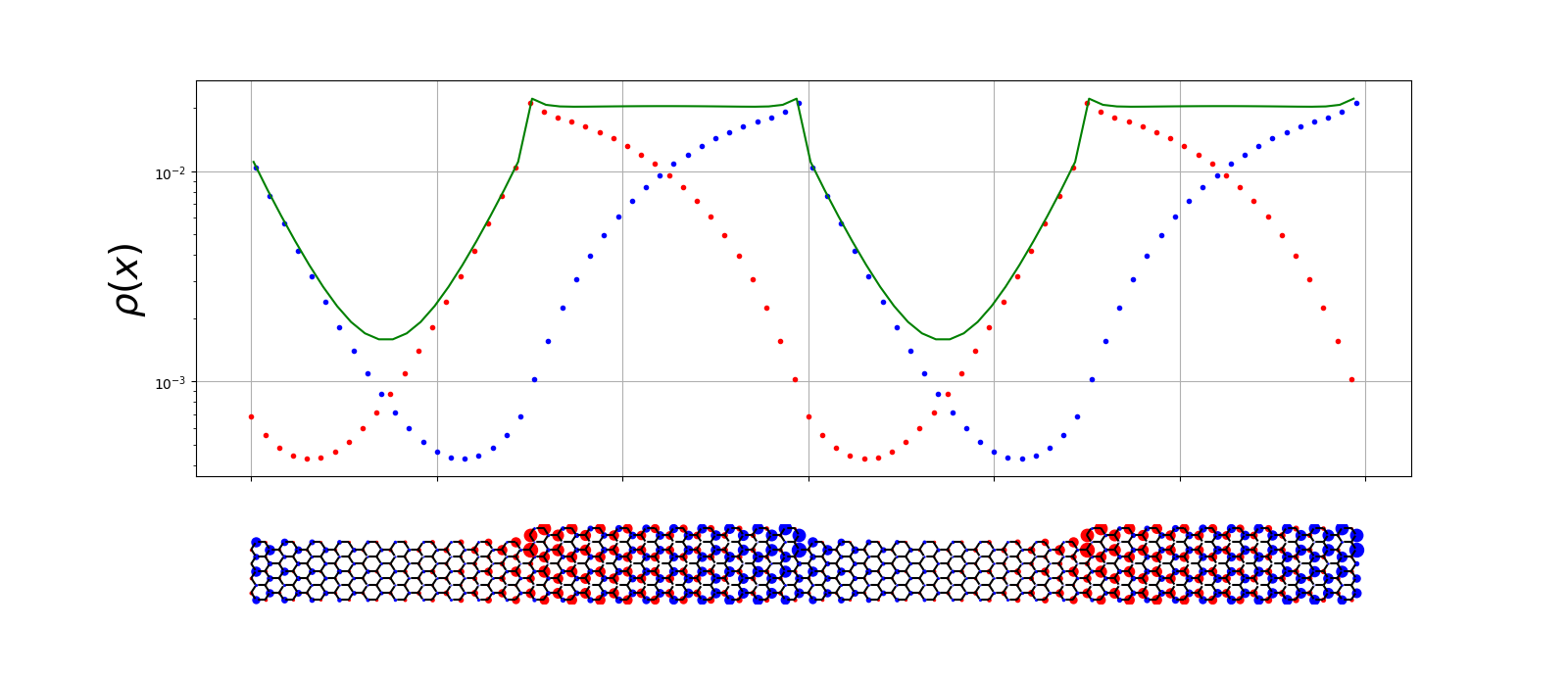}
	\caption{Lowest energy state densities of a 7/9-junction (top) and a 9/11-junction (bottom)  with
          $(m_N, m_{N+2})=(10,10)$ and aligned at the bottom rather than the center. According to Ref.~\cite{ribbon_topology}
          the 9/11-junction features a change in the topology of the respective armchairs (see \cref{tab:topology}).
          Both cases are \kil-localised since the 9- and 11-armchair sides, respectively, exhibit long range correlations.}
	\label{fig:9-11-shift}
\end{figure}

In the hybrid ribbons discussed so far the segments are aligned along their center.
In Fig.~\ref{fig:9-11-shift} we show junctions aligned along the bottom edge.
Unlike the center-aligned hybrids, the junctions of these edge-aligned hybrids do not have surplus of
one sublattice or the other and do not break the local sublattice symmetry.
This can be seen by tiling the entire hybrid ribbon with similar unit cells (closed at top and bottom as in the right panel of
Fig.~\ref{fig:7-9-junction})
so that no junction zigzag remains.
Strictly speaking, our ET breaks down in this case because no effective lattice site is generated.

Because the sublattice symmetry is locally maintained, there is no local surplus of either sublattice and we predict that no \fuj localisation is possible.
This is indeed what we observe in both cases of 7/9 and 9/11 edge-aligned junctions.
The latter has a change in topology as can be seen in Table~\ref{tab:topology} and thus poses another counterexample
to the conjecture in Ref.~\cite{ribbon_topology}.
We identify these states as another realisation of \kil-localisation; the state concentrates into the segment
with the smaller gap.

For hybrids whose segments' widths differ by more than 2 some offsets will maintain the sublattice symmetry and
some will not. We leave a detailed study of these scenarios to future work.

\section{Conclusions\label{sect:conclusion}}

When two armchair graphene nanoribbons (AGNRs) of different widths are joined symmetrically (see e.g.\ Fig.~\ref{fig:7-9-local}),
the combined system can feature a smaller band gap than either of the AGNRs and the state with energy closest to zero is
localised at the junction. Such a localisation can either be strong with correlations decaying exponentially, or weak
with a mere power law decay of correlations (typically not considered localised). We showed that the nature of this
localisation depends solely on the band gaps of the AGNRs at either side of the junction. More specifically, the
localisation is strong on one side of the junction if and only if the AGNR on this side has a non-zero gap.
This in turn is the case if and only if the ribbon is of width $N\neq 2\pmod 3$.

We discovered that, in addition to localisations on junctions, a different type of localisation is also possible,
namely a state localised within a hybrid ribbon segment as shown in Fig.~\ref{fig:kilimanjaro}. We dub the former
type of localisations `\fuj' and the latter `\kil'. \fuj\ localisations require exponential correlation decay
on both sides of the junction, therefore they are only realised by symmetric $N$/$N+2$ junctions with $N\pmod3=1$.
\kil\ localisations are much more common in that they appear in all $N$/$N+2$ hybrid AGNRs (symmetric and
non-symmetric, see Fig.~\ref{fig:9-11-shift}) without \fuj\ localisation. We observed that these results often
coincide with the topology based conjecture for \fuj\ localisations put forward in Ref.~\cite{ribbon_topology},
however, we have also identified counterexamples to the predictions from topology arguments while our description
is more fundamental and rigorous for all $N$/$N+2$ hybrid AGNRs with odd $N$.

We have derived a very simple way to predict and accurately quantify the different types of localised bound
states appearing in hybrid AGNRs. For this we reduce the initial two-dimensional tight binding problem to a
one-dimensional effective theory (ET) where the junctions of the hybrid AGNR form the sites of the 1-D lattice.
The ET also relies on a tight binding Hamiltonian~\eqref{eq:1d_Hamiltonian} which is diagonalised analytically and
the hopping amplitude between two junctions is defined solely by the ribbon connecting these junctions.
Eq.~\eqref{eq:effective_hopping} summarises this dependence. The hopping decays exponentially with ribbon length
for gapped ribbons, signifying strong localisation, and it decays as a power law for gapless ribbons resulting
in weak localisation. We have identified two parameters $\alpha,\beta$, so-called low-energy constants (LECs),
in this description that depend only on the width of the AGNR and cannot be determined other than through fitting.
We have performed these fits for odd ribbon widths up to $N\le21$ and summarised the results in table~\ref{tab:LECs}.
The same fitting procedure can easily be extended to arbitrarily broad ribbons, limited only by computing resources. Once
the LECs are determined, they can be used to predict the band gap in hybrid AGNRs, for instance yielding tuples of
respective ribbon segment lengths with the smallest gap.

Finally, we put forth an extension of our ET in the presence of Hubbard type interactions~\eqref{eq:hubbard}.
Consistent with previous findings~\cite{ribbon_junctions}, we predict the localisations to persist in the
presence of interaction and we furthermore describe the quadratic dependence of the gap on the Hubbard interaction
using an effective staggered mass term as a third LEC.

Localised \fuj-type states in armchair nanoribbons have been proposed as qubit candidates for fault-tolerant
quantum computing before~\cite{rizzo18,doi:10.1021/acsnano.1c09503,ribbon_topology,ribbon_junctions} (nicely
explained and visualised in Ref.~\cite{topo_qubits}). Their stability against perturbations make them very promising
for this application. We now add that \kil-localised states are also well suited for the same task and they
even might have some advantages, for instance that \fuj\ localisations come in alternating shapes while all
\kil\ localisations are symmetric and thus equivalent. Moreover, while localised \fuj\ states for a particular
junction type always have the same extent, \kil\ states can be smeared out over virtually arbitrary lengths,
purely governed by the length of the confining ribbon segment.


	\begin{acknowledgments}
	This work was supported in part by the Chinese Academy of Sciences (CAS) President's International Fellowship Initiative (PIFI) 
	(Grant No.\ 2018DM0034) and Volkswagen Stiftung (Grant No.\ 93562).
	It was also funded in part by the STFC Consolidated Grant ST/T000988/1 and the Deutsche Forschungsgemeinschaft (DFG,
	German Research Foundation) as part of the CRC 1639 NuMeriQS -- project
	no.\ 511713970.  Finally, we gratefully acknowledge the computing time granted by the JARA Vergabegremium and provided on the JARA Partition part of the supercomputer JURECA at Forschungszentrum Jülich.
	\end{acknowledgments}
	
	\bibliography{bibliography}
\end{document}